# Crime Hotspot Analysis and Mapping Using Geospatial Technology in Dessie City, Ethiopia

[1]Geography and Environmental study, Wollo University, Po Box 1145, Dessie, Ethiopia

Corresponding author email: hailuayene21@gmail.com

Hailu Ayene Kebede

[2]Geography and Environmental study, Wollo University, Po Box 1145, Dessie, Ethiopia

Email: mamegisandrs@gmail.com

Mohammed Motuma Assen

[3]Sociology, Wollo University, Po Box 1145, Dessie, Ethiopia

Email: meridabadisharew@gmail.com

Merid Abadi Sharew

## Abstract

Over the past few decades, crime and delinquency rates have increased drastically in many countries; nevertheless, it is important to note that crime trends can differ significantly by geographic region. This study's primary goal was to use geographic technology to map and analyze Dessie City's crime patterns. To investigate the geographic clustering of crime, the researchers used semivariogram modeling and spatial autocorrelation analysis with Moran'sI. The neighborhoods of Hote, Arada, and Segno in Dessie's central city were found to be crime-prone "hot spot" locations, as evidenced by statistically significant high Z-scores ranging from 0.037 to 4.608. On the other hand, low negative Z-scores ranging from -3.231 to -0.116 indicated "cold spot" concentrations of crime in the city's north-central sub-cities of Menafesha and Bounbouwha. With an index of 0.027492 and a Z-score of 3.297616 ($p<0.01$), the analysis overall showed a substantial positive spatial autocorrelation, suggesting a clustered pattern of crime in Dessie. The majority of crimes showed a north-south directionality, except for murder, which trended from northeast to southwest. The mean center of all crime types was found in the central Hote area. To address the complicated problem of rising crime rates in Dessie and other developing metropolitan areas, more focused and efficient enforcement techniques, and resource deployment can be informed through the knowledge acquired from the geospatial analysis.

## 1. Introduction

The rates of crime and delinquency in many nations are alarming. Over the past thirty years, police reports indicate a dramatic increase in crime across the industrialized world. The International Crime Victim Survey (ICVS), which covers six major geographic zones—Africa, Asia, Central and Eastern Europe, Latin America, and Western Europe—reveals that over 50% of urban participants experienced crime at least once between 1989 and 1996 [1]. Crime manifests in various forms, including homicide, assault, and theft [43]. Significant regional variations in crime trends exist; while violent crime has decreased in Western Europe and North America, it remains high in some Latin American and sub-Saharan African nations, as noted in the Journal of Quantitative Criminology [54]. Ethiopia, located in sub-Saharan Africa, requires further research and monitoring. Recent socioeconomic changes have contributed to rising crime rates, necessitating innovative approaches to criminal activity data management, and visualize crime patterns and allocate resources more effectively [55]. Geographic Information Systems (GIS) provide essential tools for crime mapping and analysis by correlated the quantitative and qualitative features that are suggested by its geographic characteristics [25]. These visualizations can reveal important spatial trends and inform law enforcement strategies [20].

Crime adversely affects society in both industrialized and developing nations, leading to fear, loss of life and property, decreased income, unemployment, relocation, evictions, emotional distress, and the diversion of development funds to security expenditures [5]. In Ethiopia, crime statistics from the Commission on Federal Police indicate a sharp increase in crime, impacting quality of life and community well-being. Currently, crime analysis based on spatial data has not been implemented at the national level; data collection remains manual and lacks spatial reference, which is crucial for effective crime prevention and police operations [33]. [52] notes that community policing strategies are most effective when informed by local crime data and community characteristics. The Amhara Police Commission report (2020) highlights a periodic increase in the region's crime rate, with 20,000 criminal occurrences reported in 2020—a 3.7% increase from 2019. Major crime categories included fraud, homicide, threats, theft, looting, and break-ins; however, these records lack spatial context. Understanding the relationship between individual demographics and crime is crucial for developing effective crime prevention strategies [31]. Socioeconomic factors significantly influence crime rates, emphasizing the need for comprehensive data analysis that considers community context [46]. The increase in urban crime rates is a global phenomenon influenced by various factors, including economic disparity, political instability, and rapid urbanization [18, 42]. In Ethiopia, particularly in urban centers like Dessie, recent shifts [3]. studies indicate that rising crime rates are linked to socioeconomic changes and demographic [40]. The police department in Dessie City is not yet computerized for efficient record-keeping, making it challenging to manage crime data and respond to the rising incidence of crime.

Crime has inherent spatial properties, including location, time, and process. Efficient policing relies on law enforcement agencies' ability to obtain accurate geographic information about crime-prone areas [44, 17, 32]. Previous studies, such as one focused on crime analysis and mapping in the Lideta sub-city of Addis Ababa, did not address potential criminal hotspots or significant geographic correlations [64]. Dessie City serves as a representative case for this study due to its diverse demographic composition and rapid urbanization, mirroring trends observed in other Ethiopian cities [65]. The city's significant population increase and economic shifts create a unique context for examining crime patterns [63]. This study aims to fill the gaps identified in previous research by utilizing geospatial technology to analyze and map crime hotspots in Dessie City, Ethiopia.

2. Methodology of the study

2.1. Description of the Study Area

This section provides a detailed description of Dessie City, focusing on its geographical location, demographics, climate, economy, and historical significance. Dessie City, located at an elevation of approximately 2,700 meters above sea level, is situated between longitudes 39° 35′ 36″ and 39° 40′ 45″ east and latitudes 11° 4′ 59″ and 11° 12′ 36″ north. It is approximately 400 kilometers north of Addis Ababa along the Mekelle route, in the northern section of Ethiopia. As the capital of South Wollo, Dessie is marked as 'C' on the study area map, with 'A' denoting Ethiopia and 'B' indicating the Amhara region outlined in tourmaline green. Dessie has an estimated population of around 443,570, comprising 51.21% males and 48.79% females. The population is predominantly Amhara, with significant Oromo and Tigrayan communities, contributing to a rich cultural tapestry. This diversity fosters various intercultural and religious interactions, with the two major religious groups being Muslims and Orthodox Christians. In terms of age distribution, approximately 40% are children (0-14 years), 55% are of working age (15-64 years), and roughly 5% are elderly (65 years and older) [39].

Dessie experiences a temperate climate, featuring a rainy season from June to September and a dry season from October to May. Average temperatures range from 10°C at night to 25°C during the day [62]. These climatic conditions promote agriculture, which is a vital sector of Dessie's economy. Major crops include cereals such as maize, wheat, and teff, as well as pulses like beans and chickpeas, alongside cash crops such as onions, tomatoes, and peppers. The economy of Dessie is primarily based on agriculture, with many households engaged in subsistence farming. Income levels vary widely, and a significant portion of the population lives below the poverty line. Major employment sectors include agriculture, trade, and increasingly, small-scale industries and services [23]. Access to utilities is limited, with many residents lacking reliable electricity and clean water, impacting overall living conditions. The literacy rate in Dessie is approximately 70%, with significant disparities between urban and rural areas [24]. Primary education enrolment is relatively high, but dropout rates increase in higher grades due to economic factors. Access to secondary education is improving, yet few students proceed to tertiary education [38 ,57].

Historically, Dessie has been an important commercial hub since the late 19th century, serving as a strategic point during various political and military events in Ethiopia. Notable events include its role during the Italian invasion in the 1930s and its significance as a site of resistance during the Derg regime [29]. The city's unique social dynamics can influence crime patterns, as noted by the Central Statistical Agency of Ethiopia [15]. High youth populations, combined with socioeconomic challenges, correlate with increased crime rates, particularly in areas with limited economic opportunities. Moreover, factors such as poverty and lack of education contribute to crime prevalence [6].

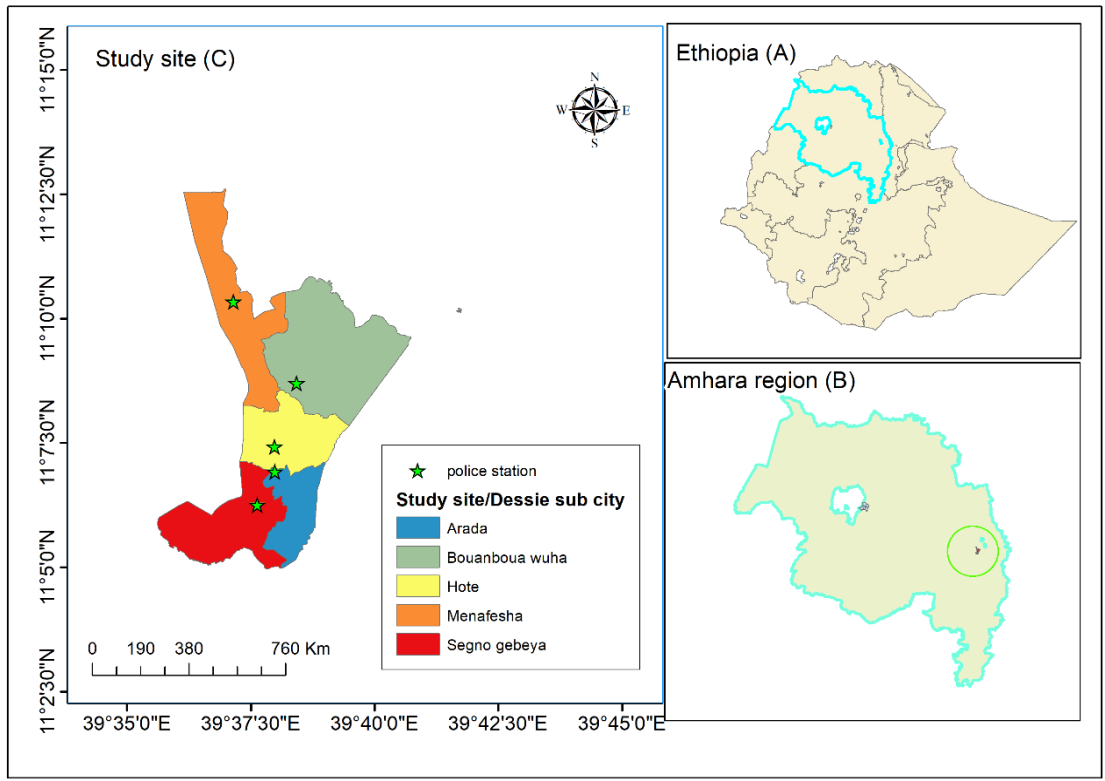

*Figure 1. Location Map of the study area*

2.2. Research Design

Substantially, both quantitative and qualitative research strategies were used in this study, since software and algorithms were utilized to process quantitative field data to produce a map and numerical results. The qualitative approach was utilized by providing a detailed narration to the software outcomes. The qualitative conclusions have since been added to and supported by the quantitative findings. The concurrent mixed research strategy is used to address this mutually beneficial utilization of the methods. This strategy has the benefit of enabling the simultaneous collection and analysis of both forms of data for reciprocal complementarity in the use of qualitative and quantitative research approaches.

2.3. Data Types and Sources

Primary data was collected using a Garmin device 72 hours to gather GPS point data for police stations and commonly occurring crime scenes. The pertinent characteristics of the crimes, such as the type of crime, age of the victims, number of victims, frequency, and time of crime occurrence in each sub-city from 2018 to 2021, were obtained through interviews with the corresponding sub-city police stations, as well as from their documents and reports.

Additionally, the researchers incorporated secondary data from books, publications, and journal articles to strengthen the paper.

**2.4. Methods of data analysis**

This study utilized the recently established optimized hotspot spatial statistics method to identify statistically significant crime hotspots and cold zones. Hotspot locations are determined based on Z- and P-score values. A characteristic is considered spatially concentrated at high values when its Z-score is high and its P-value is low. Conversely, a low negative Z-score with a low P-value indicates that low values are spatially concentrated, with the strength of clustering depending on the variations in the Z-score. When the Z-score is around zero, there is no identifiable regional grouping [21]. The quantitative methods employed in this study include semivariogram modeling, which measures how crime incidents change over distance. This technique represents the degree of similarity to check between crime events as a function of distance, allowing for the identification of spatial patterns [19]. Additionally, spatial autocorrelation analysis using Moran's I is applied; this widely used statistic assesses whether crime incidents are randomly distributed or clustered in specific areas. It quantifies the degree to which a crime occurrence at one location is similar to that at nearby locations [41]. This method is particularly useful for understanding overall crime patterns in Dessie City, as it provides a measure of spatial autocorrelation. Moreover, the optimized hotspot spatial statistics technique, utilizing tools like Getis-Ord Gi*, identifies crime hotspots and cold spots by assessing the concentration of crime incidents in specific areas. It calculates z-scores and p-values to determine the significance of these clusters. For qualitative data, interviews were conducted with local police officials to gather insights on crime trends and characteristics that may not be captured through quantitative data alone. This qualitative component provides context to the statistical findings, helping to understand the social and environmental factors contributing to crime in specific neighborhoods. All spatial analyses were conducted using ArcGIS Pro, employing tools such as Hot Spot Analysis and Inverse Distance Weighting to assess crime distribution and identify hotspots. Local indicators of spatial association (LISA) are also essential for identifying clusters of crime, enabling researchers to understand spatial patterns in crime data [4]. In further research, statistical analysis, standard deviation ellipses, mean center, directed distributions, and mean deviation ellipses were employed.

*The mean center is given as:* $\bar{x} = \frac{\sum_{i=1}^{n} x_i}{n}, \quad \bar{y} = \frac{\sum_{i=1}^{n} y_i}{n}$ ------------------------------(1)

Where $x_i$ and $y_i$ are the coordinates for features i, and n is equal to the total number of features.

The standard deviation ellipse is given as: ------------------------------------------------------- (2)

$$C = \begin{pmatrix} var(x) & cov(x,y) \\ cov(x,y) & var(y) \end{pmatrix} = \frac{1}{2}\begin{pmatrix} \sum_{i=1}^{n}\tilde{x}^2 & \sum_{i=1}^{n}\tilde{y}_i\tilde{y}_i \\ \sum_{i=1}^{n}\tilde{x}_i\tilde{y}_i & \sum_{i=1}^{n}\tilde{y}^2_i \end{pmatrix}, \quad \text{Where}$$

$$var(x) = \frac{1}{n}\sum_{i=1}^{n}(x_i - \tilde{x})^2 = \frac{1}{n}\sum_{i=1}^{n}\tilde{x}_i^2 \text{ and}, cov(x,y) = \frac{1}{n}(x_i - \tilde{x})(y_i - \tilde{y}) = \frac{1}{n}\sum_{i=1}^{n}\tilde{x}_i\tilde{y}_i$$

$$cov(x,y) = \frac{1}{n}(x_i - \tilde{x})(y_i - \tilde{y}) = \frac{1}{n}\sum_{i=1}^{n}\tilde{x}_i\tilde{y}_i, \text{ and } var(y) = \frac{1}{n}\sum_{i=1}^{n}(y_i - \tilde{y})^2 = \frac{1}{n}\sum_{i=1}^{n}\tilde{y}_i^2$$

In this case, n is the total number of features, x, and y are the coordinates for feature I, $(\tilde{x}, \tilde{y})$ is the Mean Centre for the features, and the eigenvalues and eigenvectors of the sample covariate matrix are used to represent the matrix after it is factored into a standard form. Following that, the x- and y-axis standard deviations are: -

$$\sigma 1,2 = \left(\frac{\left(\sum_{i=1}^{n}\tilde{x}_i^2 + \sum_{i=1}^{n}\tilde{y}_i^2\right) \pm \sqrt{(\sum_{i=1}^{n}\tilde{x}_i^2 - \sum_{i=1}^{n}\tilde{y}_i^2)^2 + 4(\sum_{i=1}^{n}\tilde{x}_i\tilde{y}_i)^2}}{2n}\right)^{1/2}$$

Spatial interpolation was utilized to identify potential suspect crime locations using the inverse distance weighting method. In line with [36], geo-statistical analyst tools provide (Cokriging, Kriging, Kernel Kriging, and Inverse Distance Weighting (IDW)) that enable the best possible prediction by analyzing the correlations among all sample points and creating a continuous surface of crime concentration, standard errors (predictive uncertainty), and probabilities that critical values are exceeded. This research tool establishes a connection between geostatistics and GIS.

IDW was given as: $\quad Z_j = \dfrac{\sum_i \frac{Z_i}{d_{ij}^n}}{\sum_i \frac{1}{d_{ij}^n}}, n >$ ----------------------------------------------------------(3)

Where,

$Z_i$ is the value of the known point

$dij$ is the distance to a known point

$Zj$ is the unknown point

$n$ is a user-selected exponent

The Geary's C and Moran's I tests are two methods for evaluating spatial autocorrelation. Moran's I is one method that is commonly used to search for evidence of crime incidence clustering [14]. Similar to that, Moran's I test statistic was used in this work to quantify the spatial autocorrelation of feature locations and their attribute values. This determines whether the result is random, distributed, or clustered and explains how the values relate to the feature locations. If the Z-intercept is positive, it shows that the values tend to cluster. A more thorough investigation of spatial autocorrelation was also conducted using the semivariogram.

The Moran's I statistics for spatial autocorrelation is given as: -------------------------------(4)

$$I = \frac{n \sum_{i=1}^{n} \sum_{j=1}^{n} w_{i,j} z_i z_j}{S_o \quad \sum_{i=1}^{n} z_i^2} \quad \text{-----------(4.1)}$$

$Where, z_i$ Is the deviation of an attribute for feature I from its mean $(x_i - \tilde{x})$, $w_{ij}$ is the spatial weight between features i and j, n is equal to the total number of features and $s_o$ is the aggregate of all spatial weight: $s_o = \sum_{i=1}^{n} \sum_{i=1}^{n} w_{i,j}$ -------------------------------(4.2)

The $z_i$ score for statistics is computed as:

$$z_i = \frac{I - E[I]}{\sqrt{V[I]}} \quad \text{-------------(4.3)}$$

Where, $E[I] = \frac{-1}{(n-1)}$ -------------------------------(4.4)

$V[I] = E[I^2] - E[I]^2$ -------------------------------(4.5)

The semivariogram is defined as: -------------------------------- (5)

$\gamma(s_i s_j) = \frac{1}{2} var(Z(s_i) - Z((s_j)))$, where var the variance

You would anticipate that two places, $s_i$ and $s_j$, that are close to one another in terms of the distance measure of $d(s_i, s_j)$ will be similar, therefore the difference in their values, $Z(s_j) - Z(s_i)$, will be minimal. The gap $Z(s_i) - Z(s_j)$ will increase as $s_i$ and $s_j$ Grow apart because they become less similar to one another.

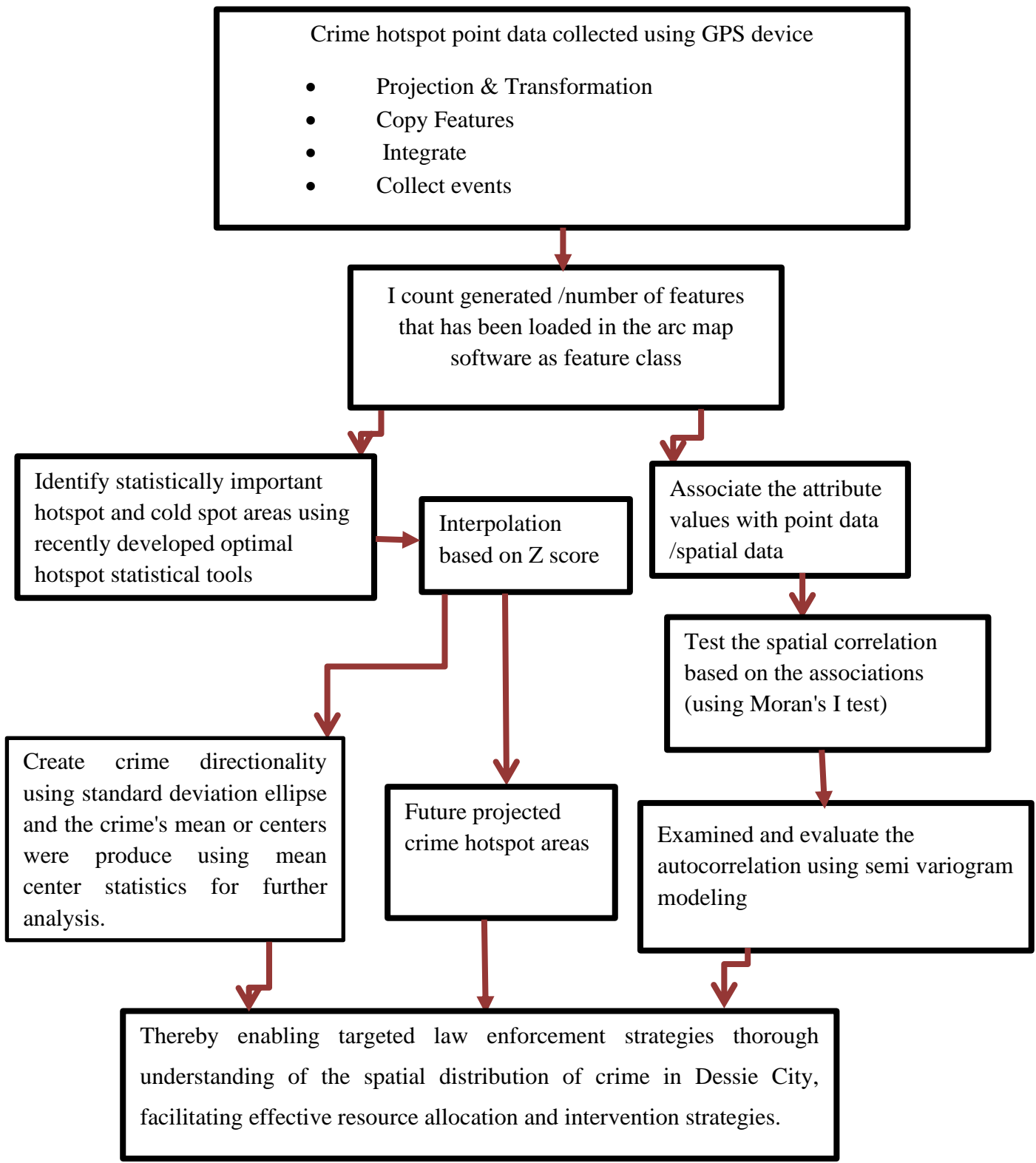

Figure 2. Flow Chart of the Study

The flowchart in Figure 2 illustrates the methodology employed in the study of crime patterns in Dessie City. It begins with the collection of primary data, which includes GPS point data from local police stations detailing crime incidents and their characteristics from 2018 to 2021. This primary data is supplemented by secondary sources to provide a comprehensive context. The study utilizes both quantitative and qualitative research methods, integrating statistical analyses with insights gained from interviews with police officials. Key analytical techniques include semivariogram modeling and spatial autocorrelation analysis using Moran's I. Additionally, standard deviation ellipse and mean center statistics are employed to identify the crime mean center and directionality.To identify potential future crime points, the study uses Inverse Distance Weighted (IDW) interpolation, along with optimal hotspot statistical tools to pinpoint crime hotspots and cold spots. The flowchart visually represents the sequential steps taken from data collection through analysis, culminating in the identification of statistically significant crime patterns. This structured approach enables targeted law enforcement strategies and ensures a thorough understanding of the spatial distribution of crime in Dessie City, facilitating effective resource allocation and intervention strategies.

## 3. Result

Understanding the different kinds of crimes and their spatial locations is crucial. Based on interviews with police professionals, the study recorded 810 spatially scattered incidents from 2018 to 2021, categorized as follows: theft (660 incidents), murder (147), intimidation (315), and other crimes. This detailed breakdown is essential for understanding the temporal and spatial dynamics of crime in Dessie [58]. In total, there were 4,462 events connected to these spatial characteristics during this period, with the following annual breakdown: 1,122 offenses in 2018, 1,106 in 2019, 1,107 in 2020, and 1,127 in 2021. Figure 3 illustrates that high crime incidents peaked in 2018, declined in 2019, and then experienced a slight increase in 2020 and 2021.

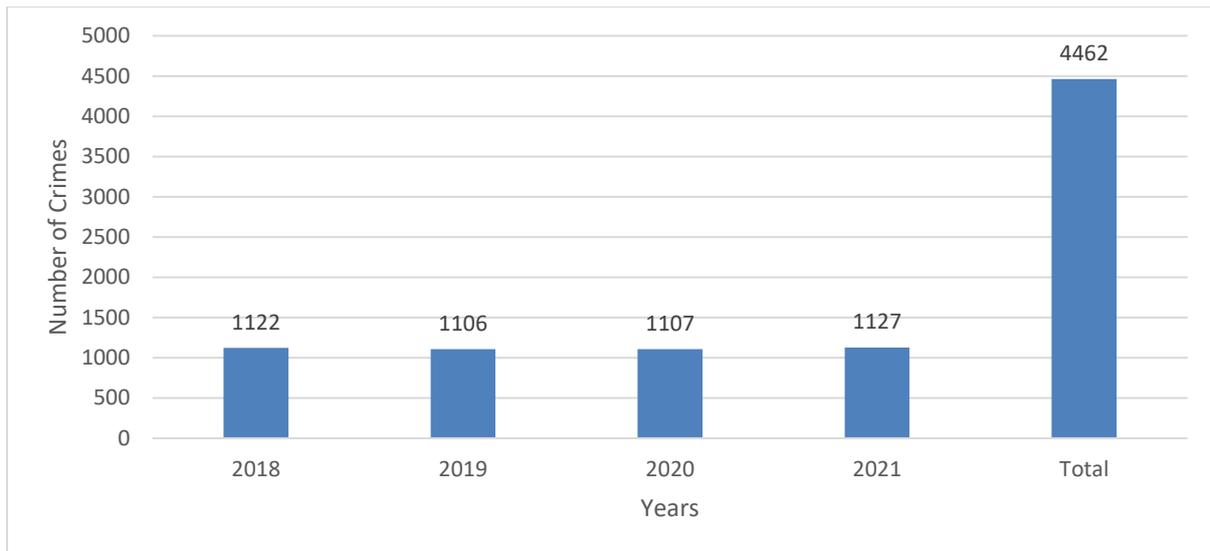

Figure 3. Observed crime from 2018 – 2021 in the study area

The identified criminal features include theft, intimidation, looting/beating, duping, burglary, and murder, as illustrated in Figure 4a. The distribution of these incidents across suburbs is as follows: 197 in Arada, 191 in Hote, 181 in Segno Gebeya, 150 in Bounbouwuha, and 91 in Menafesha. The primary locations of these crimes were vacant lots, commercial centers, notable streetside taxi fermatas, single-family residences, hotels, and market centers. Murder and other violent crimes predominantly occurred in vacant areas near the "Hote" fields. Key taxi fermatas included Arada, Paisa, Segno, Meneharia, Shel, Melaku, Menafesha, Bounbouwuha, Dandiboru, and Tekuam. Significant market centers were Segno and Robit Gebeya, along with commercial centers like Arada and Jumuruk. The frequency of crimes varied from 1 (lowest) to 28 (highest), with theft being the most common and murder the least. Based on natural breaks (Jenks) in the classification system, the range of 1 to 5 crimes is shaded in light green. Theft accounted for 660 incidents, followed by looting/beating (447), deceit (398), intimidation (315), and murder (147). Additionally, there were six to ten incidents of murder, highlighted in yellow. The most apparent crime was theft (999 incidents), followed by burglaries (45), duping (58), intimidation (483), and looting/beating (88). In the range of 11 to 16 incidents, theft (235) was the most frequent offense, colored faintly red, while intimidation (188) was the second most prevalent. In the range of 17 to 28 incidents, tinted red, the most commonly reported offenses were theft (216), followed by intimidation (156), as shown in Figure 4b.

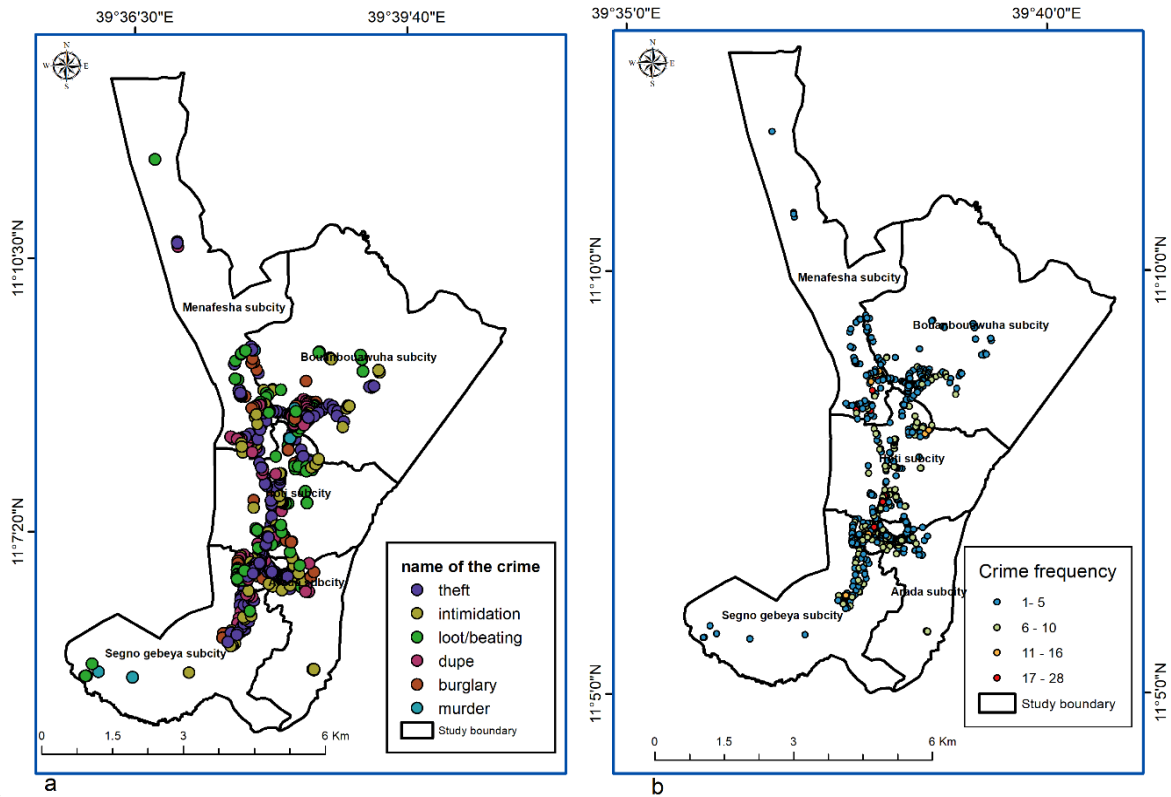

Figure 4. Distribution(a) and Frequency(b) of Crime Incidents in Dessie City 2018-2021

In temporal crime hotspots, determining the time of criminal acts is just as important as figuring out where they occur. The results of this poll showed that theft was the most typical sort of crime, accounting for 47.29 percent of all offenses. 43.14% of the events took place during the day, and 4.15% during the night. Thirteen percent were nights and twenty-seven percent were days, with anxiety accounting for twenty-five percent of them. Beats and theft accounted for 12.69% of these occurrences; 6.91% of them happened during the day and 5.09% at night. During the day, 10.22%, 7.15%, and 3.07% of people were tricked by this. Among the 4.30% that reported daytime burglaries, 2.24 percent and 2.06 percent, respectively, were from this group. Murders, which made up 0.61% of all cases, occurred both during the day and at night, as Figure 5(a) illustrates. The bulk of crimes, or 42.14% of the total, are considered thefts, as per a map showing the dates of criminal events. This suggests that larceny crimes happened more often during the day than at night, although intimidation offenses happened 12.57 the percentage of the time during the day and 12.44 proportion of the time at night, respectively. We can nevertheless conclude that intimidation occurs just as frequently during the day as it does at night, even though daytime crime rates were noticeably higher. Daytime crime was more common than night-time crime in terms of theft and burglary, whereas the opposite was true for murder. Generally speaking, just over 30% of crimes happened at night and more than 70% happened during the day.

In demographic crime hotspots, residents are often victimized based on their age, sex, or other demographic characteristics as highlighted in a paper by the Dessie City Police Department. [53 56 35], argued that individuals become victims of crime due to these demographic factors. In Ethiopia, individuals are categorized into three age groups: young (0–14), working age (15–64), and old age (65 and over) [16]. Using these age groupings, a mapping exercise was conducted for this investigation, as illustrated in Figure 5b. The findings indicate that within the young age group, females constitute 2.73% of victims, while males account for 0.14%. In the working age group, females make up 16.07% of victims, and males represent 25.46%. For the old age group, the percentages are 1.28% for females and 54.32% for males. Notably, the old age group experiences the highest victimization rate, comprising 55.60% of total victims. This heightened vulnerability may stem from factors related to aging, which can make individuals more susceptible to exploitation and fraud. Additionally, the working age group constitutes 41.53% of the total cases. Collectively, these two age groups account for the vast majority of victims (97.13%), while the young age group represents a minimal 2.87%.

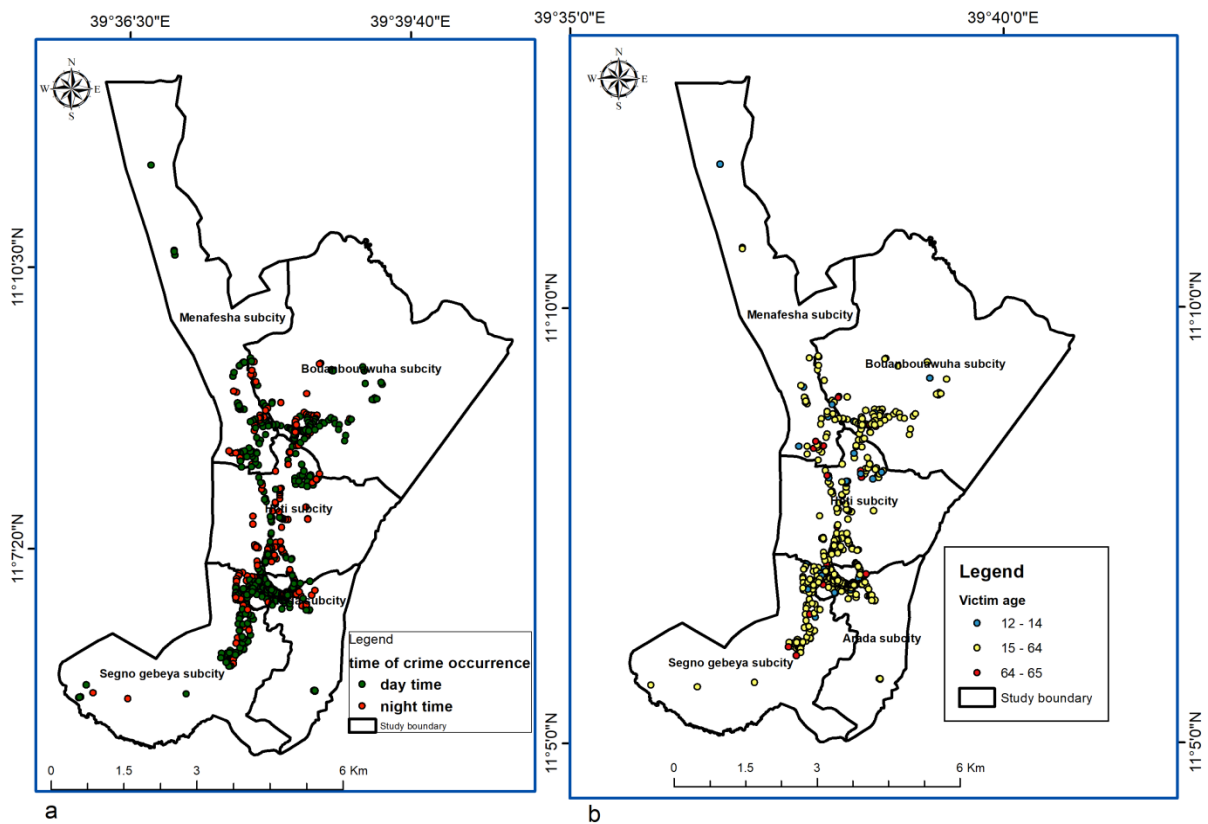

Figure 5. Temporal Distribution of Crime Incidents(a) and Victim Age Demographics(b) in Dessie City 2018-2021

The application of ArcGIS Pro Hot Spot Analysis techniques allows for the identification of spatial clusters with statistically significant high or low attribute values, particularly concerning criminal episodes. By utilizing a collection of weighted data points, such as the total number of crimes per census block, analysts can effectively pinpoint areas with elevated crime rates, referred to as hot zones. This method not only highlights clusters of high criminal activity but also identifies regions where crime rates are unexpectedly low, providing valuable insights into neighborhood characteristics or legislative measures that may deter crime [37, 45]. The Hot Spot Analysis tool calculates the Getis-Ord Gi* statistic, which produces z-scores and p-values that indicate the significance of these clusters. Statistically significant positive z-scores reveal intense clustering of high values (hot spots), while significant negative z-scores indicate clustering of low values (cold spots) [22]. In this context, areas identified as Hote, Arada, and Segno are recognized as statistically significant hotspots, suggesting a concentration of criminal activity. Conversely, the analysis of Menafesha and Buanbuawuha, despite their modest negative z-scores, indicates the presence of statistically significant cold patches, highlighting areas where crime is lower than expected. Figure 6 illustrates the spatial distribution of crime hotspots and cold spots within Dessie, emphasizing areas with statistically significant concentrations of criminal activity, which are critical for targeted interventions [8].

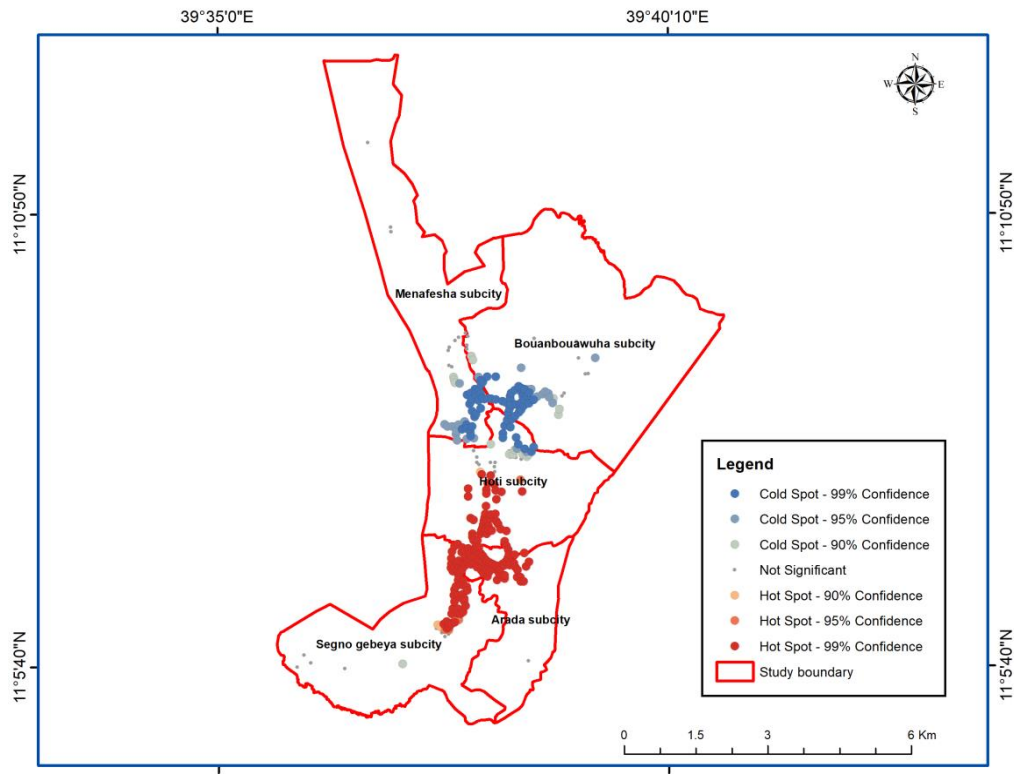

Figure 6. Map Showing Statistically Significant Crime Hotspots and Cold Spots in the Study Area

Spatial autocorrelation is a measure of how effectively nearby objects correlate with each other across a specific spatial region. When many similar values are separated by significant distances, negative autocorrelation can result. According to [27], spatial autocorrelation plays a crucial role in defining the relevance of geographic variables, particularly concerning how they influence specific items in space when a clear relationship exists among objects and spatial features. Moran's I is a statistical measure used to quantify spatial autocorrelation across a region [13]. In ArcGIS pro, the "Spatial Autocorrelation (Moran's I)" tool can effectively measure this phenomenon. Geographical units exhibit strong positive values when located near similar units, while strong negative values occur when they are near dissimilar units. A Moran's I value of 0 indicates neither autocorrelation nor independence among geographic units.

In our study, spatial crime patterns were analyzed using the Moran's I spatial autocorrelation statistical approach, which also aids in identifying hotspots and cold spots. The tool calculates the Moran's I index along with a z-score and a p-value to determine the significance of the index. Our results showed a z-score of 3.297616 (greater than 2.58) and a p-value of 0.000975 ($p < 0.01$), indicating a strong positive spatial autocorrelation with an index of 0.027492. These results suggest that the pattern was clustered with 99% confidence as shown in Figure 7a. As suggested by Caplan and his friends when spatial autocorrelation is present, a spatial lag control variable should be incorporated into the statistical model. Additionally, semi-variogram modeling was utilized to evaluate and quantify the autocorrelation, resulting in the model: 0.76801Nugget + 0.28899Stable (136.32, 1.216) as shown Figure 7b. A non-zero nugget indicates that the data exhibit spatial autocorrelation [10], the presence of a nugget (0.76801) suggests measurement errors or significant variability at very small distances. Conversely, the stable component (0.28899) reflects a considerable degree of spatial correlation at greater distances, indicating that spatial correlation eventually diminishes. The range (136.32) indicates the distance over which this spatial correlation occurs, suggesting that points within this range are associated. The nugget and stable components together create a sill, representing the overall variance at which the variogram levels off. A higher sill indicates significant geographical heterogeneity within the study area. The parameters of the semi-variogram model demonstrate a strong association between points, as implied by the correlation coefficient (0.76801), which is close to 1. Overall, these findings strongly support the existence of spatial autocorrelation within the data.

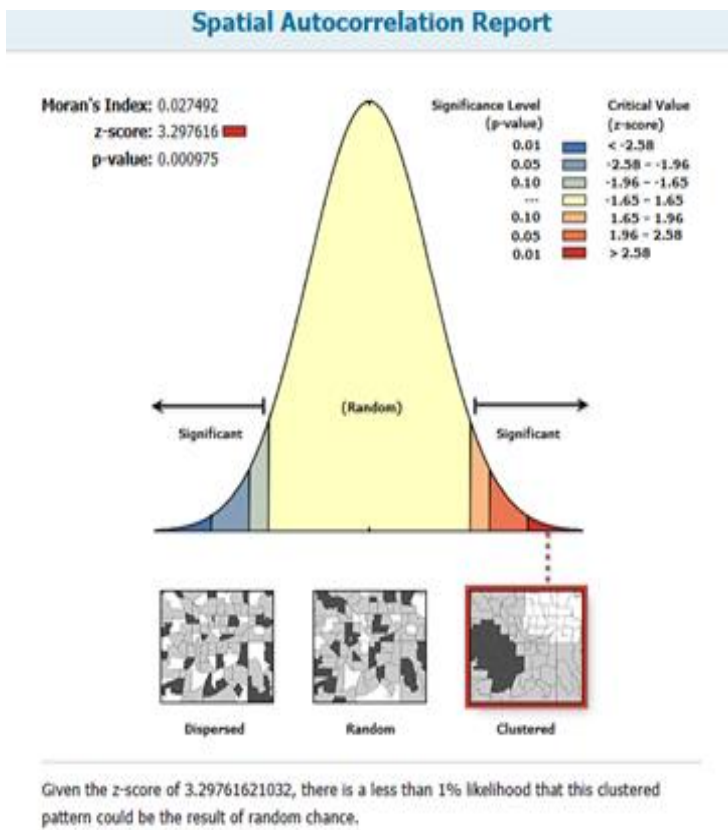 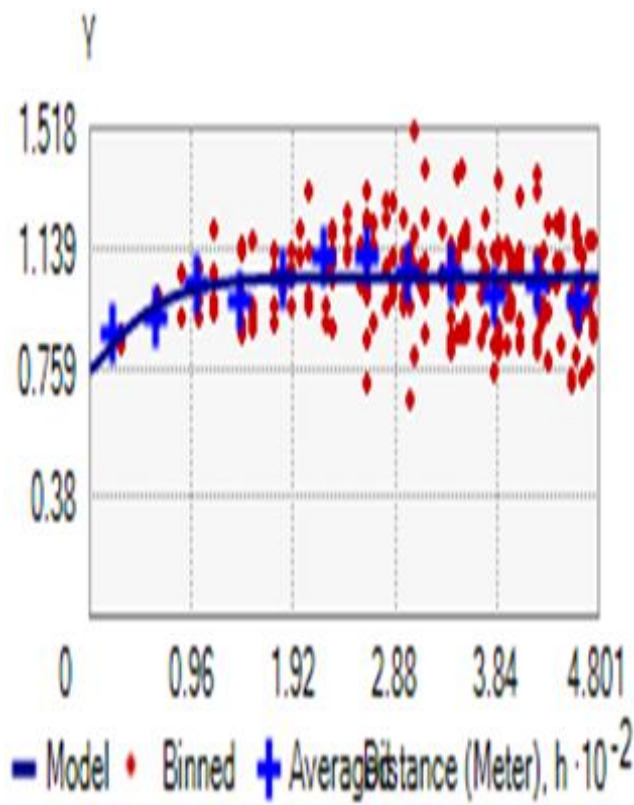

Figure7. Spatial Autocorrelation Analysis(a) and Semi variogram Modeling(b) in the Study Area

Besides crime hotspot area identification, crime prediction can be a vital asset for the police department as well as society [26]. Weighted geographical interpolation was used to generate the possible crime scene or future crime hub spots (the unidentified points) based on a mathematically distance-based algorithm that identified hub sites that were already accessible. The tool determines cell values by linearly weighing a combination set of sample points. According to [48], the inverted distance weight is named because it assigns a higher weight to locations that are closer to the predicted site than to those that are further away. Each intake point has a localized impact that gets smaller with distance, which is taken into consideration by the popular interpolation technique known as Inverse Distance Weight (IDW). As can be observed in Figure 8a, points nearer the processing cell are given a higher weight than points farthest from it. Based on the currently identified crime hotspots, future crime hotspot locations were estimated for an IDW study. A map of the research area that has been interpolated shows that the northern and southern regions/ periphery parts of the study, which are colored light red, have a lower concentration of crime than the middle parts, which are colored green and yellow include Hote, Segno gebeya, and Arada sub-city.

Due to the method's determination of the average x- and y-coordinates from the mean center, which establishes the ellipse's axis, the oval form is known as an ordinary deviational ellipse. Because of the oval, it is possible to determine if the distinctive dispersion is prolonged and, therefore, has a particular orientation. The claim put forth by [2], that oval deviations can be utilized to show dispersal and direction is similar to this. The oval's size and shape account for the degree of dispersion, while the alignment of the ellipse accounts for the direction of the crime type. Additionally, dispersion is defined by SDE as the departure from the mean of each feature location's distance from the mean center, together with the direction of that dispersion, according to [60]. The mean center, often known as the central spot, is the arithmetic mean of all feature positions. Consequently, all mean centers, or the points where the spatial feature placements in the data were balanced, were discovered in the middle, which is Hote as shown in Figure 8b, except for murder, which was situated on the southwest boundary of the sub-city. To monitor and promptly capture offenders, it is suggested that a community police station be set up in these areas. All of the study's crimes had a north-south directionality, according to the standard deviation ellipse map, except for murder, which had a north-east-south-west directionality. This may indicate that the area in question had a higher concentration of crime.

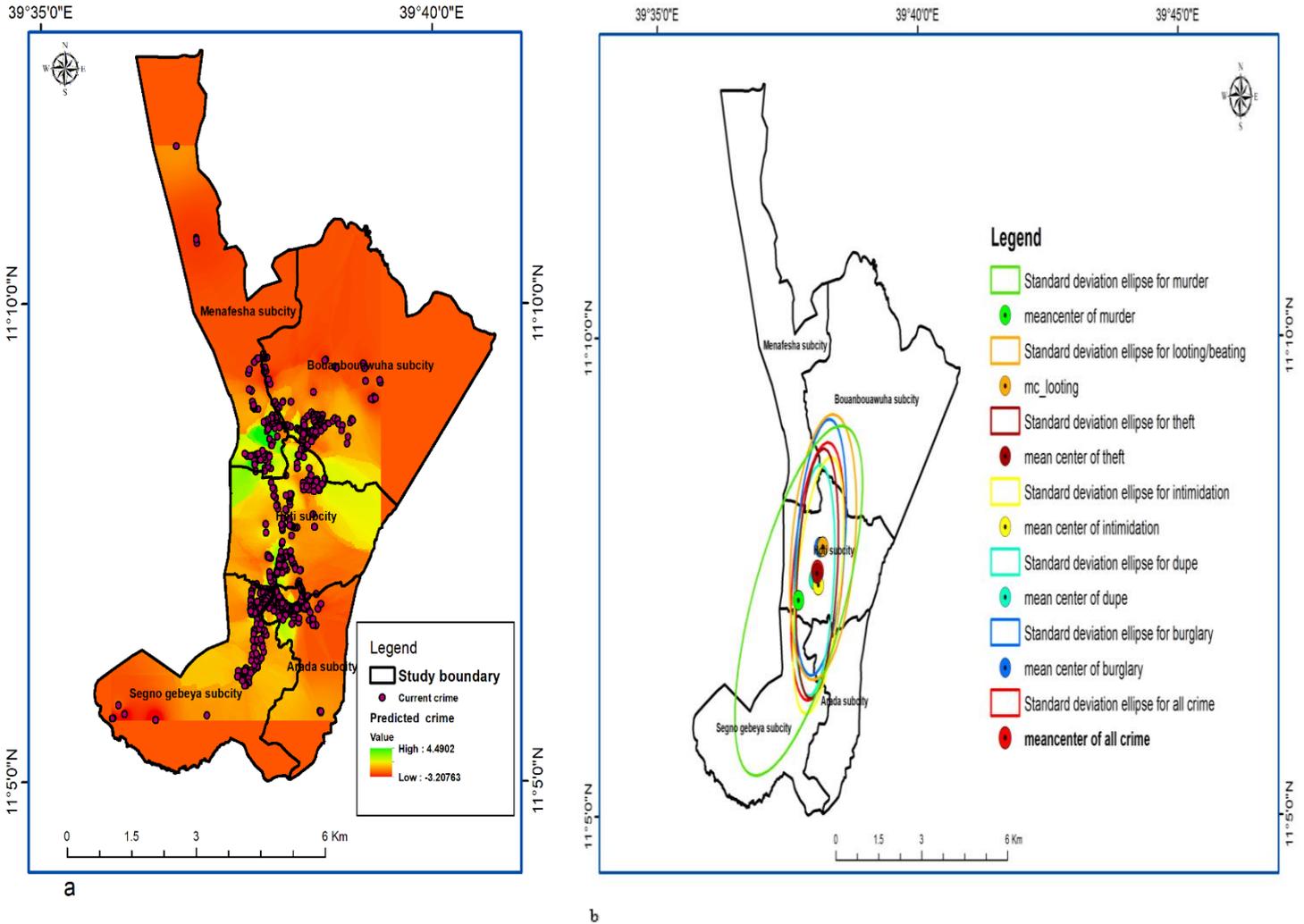

Figure 8. Interpolated Crime Hotspot Areas(a) and Standard Deviation Ellipse with Mean Center(b) in the Study Area

4. Discussion

This study utilized the Optimized Hot Spot spatial statistical tools in ArcGIS 10.7 to generate new feature outputs, including z-scores, p-values, and Gi Bin confidence levels. Various geographic statistical approaches can identify hot and cold spots, as noted by [59]. The features within the bins indicate a confidence level of 90% at a p-value of 0.01, while a Gi Bin value of zero signifies non-significant clustering. Notably, bins reflecting a 99% confidence level correspond to a p-value of 0.10, while those at a 95% confidence level correspond to a p-value of 0.05. It is important to differentiate between visually identified "hot spots" and those deemed statistically significant. For a feature to be classified as statistically significant, it must have a high value surrounded by other high-value characteristics. Similarly, cold spots consist of low-value features surrounded by other low values, indicating a pattern that exceeds random chance. The "Hot Spot Analysis" functionality in ArcGIS Pro employs rigorous statistical methods to calculate the z-score for each feature in the dataset [11]. A lower, statistically significant negative z-score indicates a stronger cluster of low values (cold spots), while a higher positive z-score signifies a more intense cluster of high values (hot spots). The study identified red-tinted high z-scores ranging from 0.037 to 4.608 in statistically significant crime hot zones, with notable features including Segno, Arada, and Hote.

The higher concentration of crime in these areas can be attributed to several factors: Hote's central location, its bus station, and open spaces contribute to its status as a crime hotspot. Additionally, the Central Business District (CBD) in Arada and the proximity of Segno to major thoroughfares also play a role. Conversely, low z-scores (ranging from -3.231 to -0.116) identified cold spots in Menafesha and Bounbouwha, highlighting areas with significantly fewer crimes. Environmental criminology integrates everyday activities, crime pattern theory, and rational choice theory [49]. This theory posits that crime patterns depend on the temporal and spatial distribution and interaction of targets, offenders, and opportunities, helping to identify areas most susceptible to criminal activity. Research indicates that approximately 5% of street segments, or "micro-places," account for over 50% of all criminal occurrences [28, 61, 51]. This study confirms that the majority of crime is concentrated in the south-central area of the city, as indicated on the crime hotspot map.

Identified crime-prone areas include market centers, open spaces, narrow streets, and intersections with high population densities, particularly among the working-age population and low-income individuals. The presence of drugs and alcohol in these areas correlates with higher crime rates. The standard deviation ellipse map further revealed that most crimes followed a north-south trajectory, except for murder, which exhibited a northeast-southwest directionality. Given that theft was the most commonly reported crime, occurring more frequently during the day, these findings align with previous research on crime patterns in Dessie and similar urban settings [30]. Other studies have noted higher crime rates in densely populated areas, such as bus stations and main thoroughfares [47, 34]. The identification of Hote, Arada, and Segno as crime hotspots aligns with established theories of environmental criminology, which suggest that crime often concentrates in specific locations due to factors such as opportunity and social disorganization [7, 50].

This underscores the necessity for tailored law enforcement strategies in these areas to effectively combat crime. The results of this study highlight the importance of using geospatial analysis in crime prevention strategies. Policymakers can leverage these insights to deploy resources more effectively, focusing on identified hotspots to reduce crime rates [12]. [9] note, policing crime hotspots through targeted interventions can lead to significant reductions in crime rates, particularly in urban areas.

**Declaration of competing interest**

The authors have not declared any conflict of interest.

**Acknowledgment**